\documentstyle[multicol,aps,epsfig]{revtex}
\begin{document}
\title{Finite-temperature magnetism of transition metals:\\
an LDA+DMFT approach.}
\author{A. I. Lichtenstein and M. I. Katsnelson$^*$}
\address{University of Nijmegen, NL-6525 ED Nijmegen, The Netherlands}
\author{G. Kotliar}
\address{Serin Physics Laboratory, Rutgers University, Piscataway, New Jersey 08855}
\date{\today}
\maketitle

\begin{abstract}
We present an {\it {ab initio}} quantum theory of the finite temperature
magnetism of iron and nickel. A recently developed technique which combines  
dynamical mean-field theory with realistic electronic structure methods,
successfully describes the  many-body features of the one electron spectra
and the observed  magnetic moments  below and  above the Curie temperature.
\end{abstract}

\pacs{75.30.-m, 71.15.-m, 71.27.+a}

\begin{multicols}{2}

The theory of itinerant electron ferromagnetism is one of the central
problems in condensed matter physics (for reviews see \cite{Vons,Vollhardt}). 
There is a need for a  first principles approach
which is able to describe ground state  and thermodynamical
quantities, as well as the one particle spectral properties
of itinerant magnets.  These quantities 
are currently being probed in spin polarized tunneling
as well as spin polarized photoemission experiments with a view
to possible applications as spin valves\cite{tunnel}.

Iron  and nickel are the oldest and
experimentally best studied prototypical systems, and 
serve as benchmarks for  electronic structure methods.   
At very low temperatures, a band like description
of Fe and Ni has been very successful.
The density functional theory (DFT) 
in the local density approximations (LDA) gives a quantitatively accurate
description of several ground state  properties of these materials such
as the ordered magnetic moment and the spin wave stiffness \cite{RPA} as
calculated from the spin wave dispersion.

While density functional theory, can in principle provide a rigorous
description of the finite temperature thermodynamic properties, 
at present there is no accurate practical implemention available. 
As a result the finite temperature properties of magnetic materials
are estimated following a simple suggestion \cite{LKAG}, whereby constrained
DFT at T=0 is used to extract exchange constants for a 
{\it classical} Heisenberg model, which in turn is solved 
using  approximation methods (e.g. RPA, mean field ) from
classical statistical mechanics of spin systems \cite{LKAG,RJ,Eschrig,Antropov}.
The most recent implementation of this approach gives good values for
the transition temperature of iron but not of nickel \cite{kudrnovski}. 
While these localized spin models give, by construction,
at high temperatures a Curie-Weiss like magnetic susceptibility,
as observed experimentally in Fe and Ni,  
they encountered difficulties in predicting the correct 
values of  the Curie constants\cite{gyorffy}.

It has been recognized for a long time, that to describe the finite
temperature aspects of itinerant electron magnets, one needs a formalism
that takes into account the existence of local magnetic moments 
present above $T_{C}$ \cite{LMM,moriya}.
This one of  the central  problems in the physics of strongly
correlated electron systems, which forces us to reconcile      
the dual character  of the electron, which as a particle  requires a  
real space  description and as a  wave requires  a momentum
space description  in a unified framework.
A very successful method satisfying these
requirements, the dynamical mean-field theory (DMFT) \cite{review}, 
has been recently developed.  This many-body scheme can be
combined with standard LDA band structure calculations to include the
effects of a  realistic band structure \cite{Anisimov,LK98}. Such
``LDA+DMFT'' approach has been successfully applied for the computations of
electronic structure and spin-wave spectrum of iron \cite{KL99,KL00}.
Nevertheless the most difficult and interesting finite-temperature effects
were not considered previously and are the subject of this letter.

Here we present realistic LDA+DMFT calculations of finite temperature
magnetic properties of iron and nickel. A numerically exact
Quantum Monte-Carlo (QMC) scheme is used for the solution the DMFT equations.
We find that a consistent first principles  description
of the magnetic properties and of the one electron spectra
of iron and nickel is possible, within an
approach that makes only two essential approximations: the locality
of the electron  self energy and of the   particle  hole irreducible
vertex\cite{review}.

There have been numerous efforts to construct a many body theory of
these materials,
for example using the T-matrix theory \cite{Liebsch},
a self-consistent moment method \cite{Nolting},
local 3-body equations \cite{Igarashi,Manghi} and the
GW\cite{Ferdi} approximations. The LDA+DMFT approach  goes  beyond
these works  in the treatment of a realistic  
orbitally degenerate band structure, and in the treatment
of the many body interactions. 

We start with the LDA Hamiltonian in the tight-binding orthogonal LMTO
representation
$H_{mm^{\prime }}^{LDA}({\bf k})$
\cite{OKA}, where $m$ describes the orbital
basis set containing  $3d$-, $4s$-  and $4p$- states.
${\bf k}$ runs over
the Brillouin zone (BZ). The interactions
are parameterized by a matrix of  screened local
Coulomb interactions $U_{mm^{\prime }}$ and
and a matrix of   exchange constants $J_{mm^{\prime }}$,
which are   expressed in terms of  
two screened Hubbard  parameters U and J, describing
the average Coulomb repulsion  and the  interatomic
ferromagnetic exchange respectively.
We use the values 
$U=2,3$ $(3,0)$ eV for Fe (Ni) and 
the same value of  the interatomic exchange,
$J=0.9$ eV
for both Fe and Ni,
a result of  constrained LDA calculations
\cite{Coulomb,Anisimov,LK98}.

Dynamical mean-field theory maps the many-body  system
onto a  multi-orbital quantum impurity, i.e. a set of 
local degrees of freedom in a bath  described by the Weiss
field function ${\cal G}$. The impurity action
(here $n_{m\sigma }={c}_{m\sigma }^{+}c_{m\sigma }$
and  ${\bf c}(\tau )=[c_{m\sigma }(\tau )]$ is a vector of  Grassman
variables) is given by:
\begin{eqnarray}
&&S_{eff} =\int_{0}^{\beta }d\tau \int_{0}^{\beta }d\tau ^{\prime }Tr[{\bf c}%
^{+}(\tau ){\mathbf {\cal G}}^{-1}(\tau ,\tau ^{\prime }){\bf c}(\tau ^{\prime
})]+   \label{path} \\
&&\frac{1}{2}\sum_{m,m^{\prime },\sigma }
\int_{0}^{\beta }d\tau [U_{mm^{\prime }}n_{\sigma
}^{m}n_{-\sigma }^{m^{\prime }}+(U_{mm^{\prime }}-J_{mm^{\prime }})n_{\sigma
}^{m}n_{\sigma }^{m^{\prime }}]  \nonumber
\end{eqnarray}
It describes the spin, orbital, energy and temperature dependent interactions
of particular magnetic $3d$-atom with the rest of the crystal and is 
 used to compute the  local Greens function matrix: 
\begin{equation}
{\bf G}_{\sigma }(\tau -\tau ^{\prime })=-\frac{1}{Z}\int D[{\bf c},{\bf c}%
^{+}]e^{-S_{eff}}{\bf c}(\tau ){\bf c}^{+}(\tau ^{\prime })
\label{pathint}
\end{equation}
($Z$ is the partition function)
and the impurity self energy
${\mathbf {\cal G}}_{\sigma }^{-1}(\omega _{n})-{\bf G}%
_{\sigma }^{-1}(\omega _{n})={\bf \Sigma }_{\sigma }(\omega _{n})$ .

The Weiss field function is required to obey the self consistency
condition  \cite{Anisimov,LK98}, which restores translational invariance
to the impurity model description:
\begin{equation}
{\bf G}_{\sigma }(\omega _{n})=\sum_{{\bf k}}[(i\omega _{n}+\mu ){\bf 1}-%
{\bf H}^{LDA}({\bf k})-{\bf \Sigma }_{\sigma }^{dc}(\omega _{n})]^{-1}
\label{BZI}
\end{equation}
 $\mu $ is the chemical potential defined self-consistently through the
total number of electrons, $\omega _{n}=(2n+1)\pi T$ are the Matsubara
frequencies for temperature $T\equiv \beta ^{-1}$ ($n=0,\pm 1,...$) and $%
\sigma $ is the spin index. The local  matrix ${\bf \Sigma }%
_{\sigma }^{dc}$  is the sum of two terms, the impurity self energy
and  a so-called ``double counting ''  correction, $E_{dc}$ which is
meant to substract 
 the average electron-electron interactions  already
included in the LDA Hamiltonian. For metallic systems we propose the general
form of dc-correction: ${\bf \Sigma }_{\sigma }^{dc}\left( \omega \right) =%
{\bf \Sigma }_{\sigma }\left( \omega \right) -\frac{1}{2}Tr_{\sigma }{\bf %
\Sigma }_{\sigma }\left( 0\right) $. This is motivated by  the fact that the
static part of  the correlation effects 
are already well described in the density functional theory. 
Only  the $d$-part of the self-energy is presented in our
calculations, therefore ${\bf \Sigma }_{\sigma }^{dc}=0$ for $s$- and $p$-
states as well as for non-diagonal $d-s$,$p$ contributions. In order to
describe the finite temperature ferromagnetism of transition metals we use
the {\em non} spin-polarized LDA Hamiltonian ${\bf H}^{LDA}({\bf k})$
and accumulate{\em \ all} temperature-dependent spin-splittings in the
self-energy matrix ${\bf \Sigma }_{\sigma }^{dc}\left( \omega _{n}\right) $.

We use the impurity QMC scheme for  the solution of the multiband
DMFT equations \cite{Rozenberg}.  The Hirsch discrete
Hubbard-Stratonovich transformations  introduces $(2M-1)M$ auxiliary
Ising fields $S_{mm^{\prime }}^{\sigma \sigma ^{\prime }}$ where $M$ is the
orbital degeneracy of the d-states and calculate ${\bf G}_{\sigma }(\tau )$ 
by an exact integration of the fermion degrees of freedom
in the functional integral (Eq.(\ref{pathint})) \cite{review}. In order to
sample  efficiently all the spin configurations in the multi-band QMC scheme,
it is important to use  ``global'' spin-flips: $[S_{mm^{\prime
}}^{\sigma \sigma ^{\prime }}]\rightarrow \lbrack -S_{mm^{\prime }}^{-\sigma
-\sigma ^{\prime }}]$ in addition to the local moves of the auxiliary
fields. The number of QMC sweeps was of the
order of 10$^{5}$. A  parallel version of  the DMFT program was used
to sample the 45 Ising fields for $3d$-orbitals. We used 256 ${\bf %
k}$-points in the irreducible part of the  BZ
for the k integration.  10 to 20 DMFT
iterations  were sufficient to achieve convergence far from the Curie point.
Due to the cubic symmetry of the bcc-Fe and fcc-Ni lattices the
local Green function is diagonal in the basis of real spherical harmonics. 
The spectral functions for real frequencies were obtained from the QMC
data by applying the maximum entropy method \cite{MEM}.

\begin{figure}[tbp]
\centerline{\epsfig{file=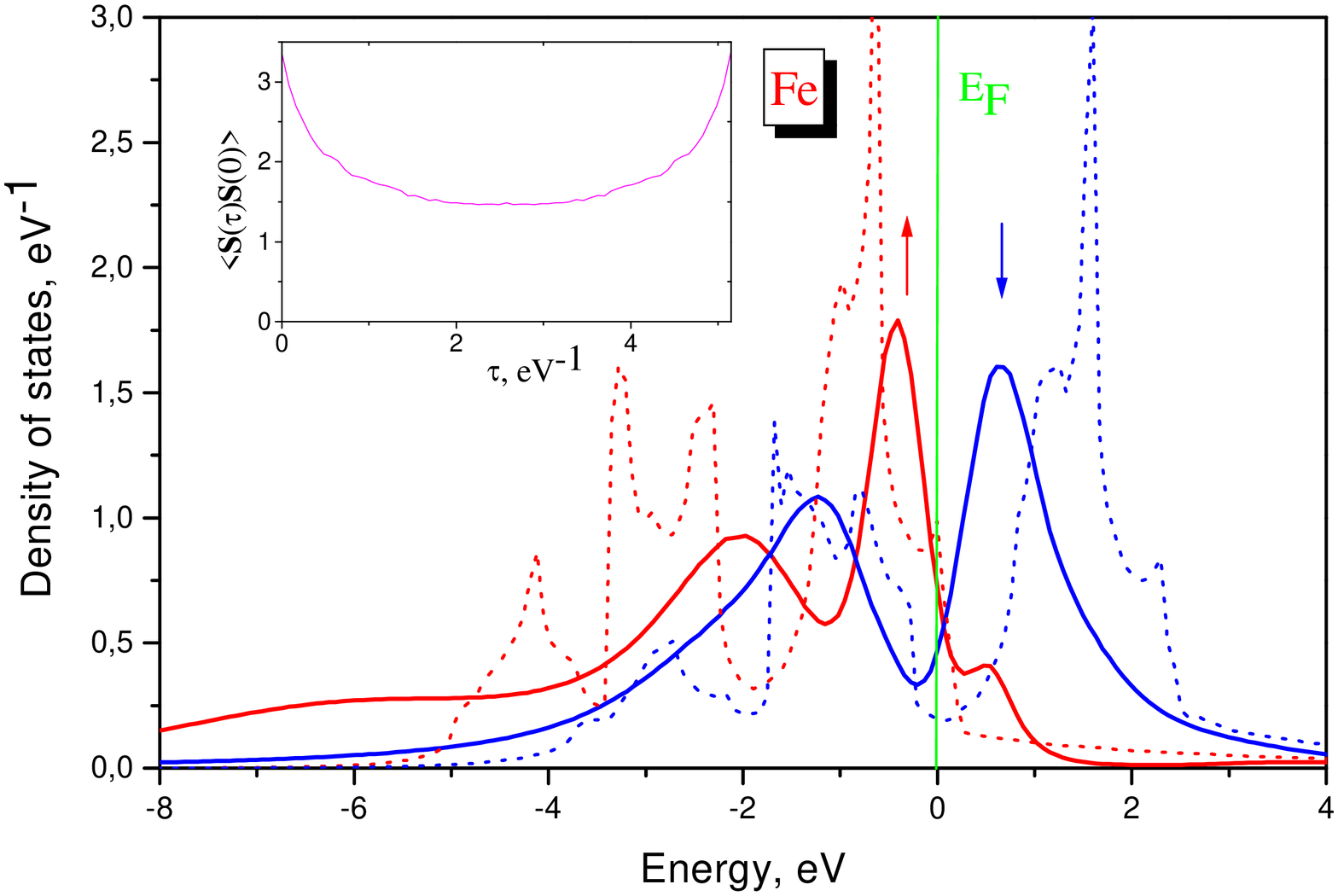, width=8cm,height=6cm,angle=0} }
\caption{LDA+DMFT results for ferromagnetic iron ($T=0.8$ $T_{C}$).
The partial densities of d-states (full lines) is
compared with the corresponding LSDA results at zero temperature (dashed
lines) for the spin-up (red lines, arrow-up) and
spin-down (blue lines, arrow-down) states. 
The insert shows the spin-spin autocorrelation
function for T=1.2 $T_{C}$. }
\label{DOSFe}
\end{figure}
Our results for the  local spectral function
for iron and nickel are shown in Figs.1 and 2,
respectively. The LDA+DMFT approach  describes  well all the qualitative
features  of the
density of states (DOS), which is especially non-trivial for nickel. Our QMC
results reproduce well the three main correlation
effects  on the one particle spectra below $T_{C}$ \cite{himpsel}:
the presence of a famous 6 eV
satellite, the 30\% narrowing of the occupied part of $d$-band and the 50\%
decrease of exchange splittings compared to the LDA results. Note that the
satellite in Ni has substantially  more spin-up contributions in agreement with
photoemission spectra\cite{himpsel}.
Correlation effects in Fe are less pronounced than in Ni,
due to its large  spin-splitting and the  characteristic bcc-structural
dip in the density of states for spin-down states near Fermi level,
which reduces the density of  states for particle hole excitations.

\begin{figure}[tbp]
\centerline{\epsfig{file=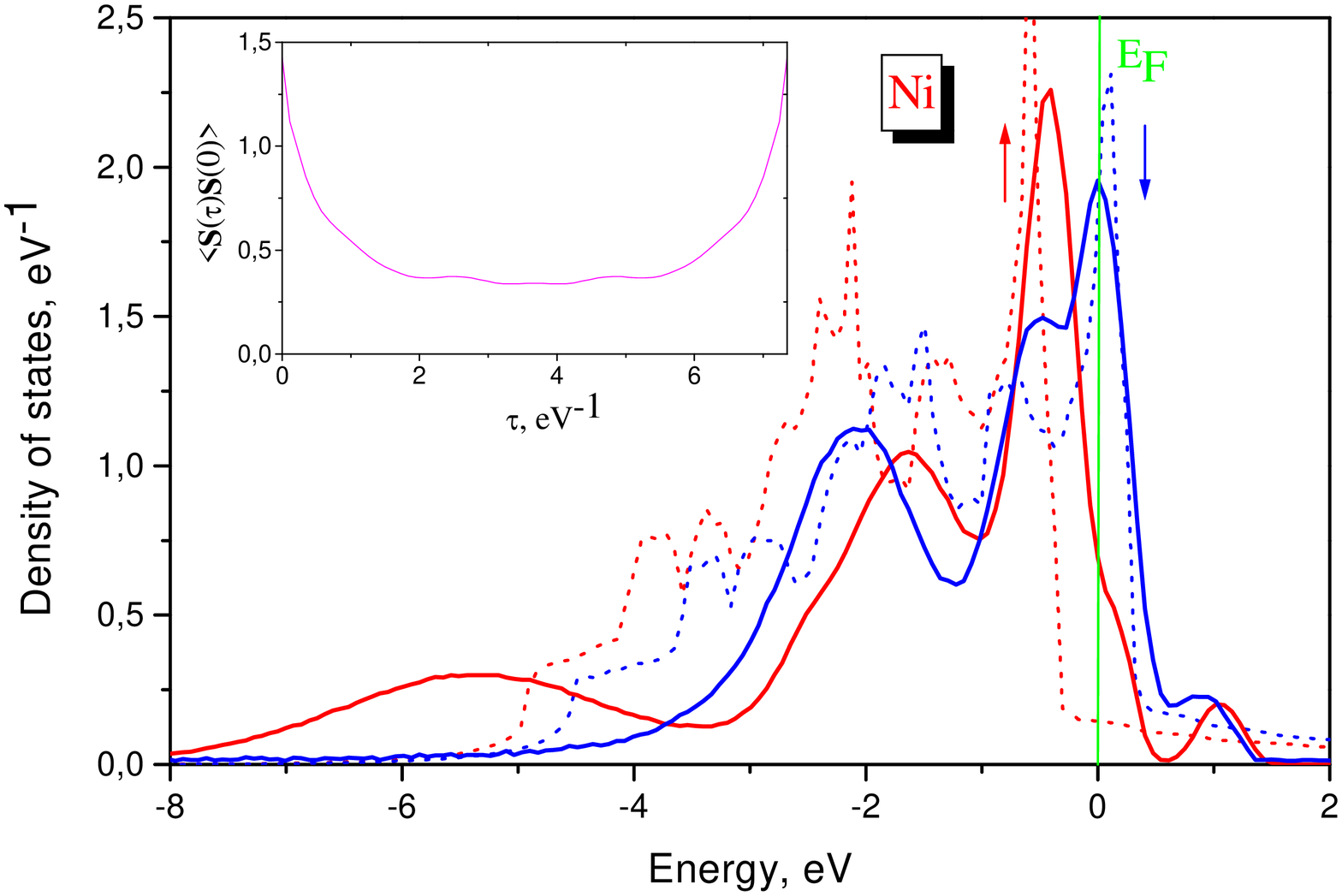, width=8cm,height=6cm,angle=0} }
\caption{ Same quantities as in  Fig.1 for ferromagnetic nickel
($T=0.9$ $T_{C}$).
The insert shows the spin-spin autocorrelation
function for T=1.8 $T_{C}$. }
\label{DOSNi}
\end{figure}
The uniform  spin susceptibility in the paramagnetic state:
\begin{equation}
{\chi _{q=0}}={\frac{dM}{dH}}  \label{qzero}
\end{equation}
was extracted from
the QMC simulations by measuring the induced magnetic moment in a small
external magnetic field. It includes  
the polarization of  the impurity Weiss field by the external field 
\cite{review}. 

The dynamical mean field results accounts for
the observed Curie -Weiss law which is observed experimentally in Fe and Ni.
As the temperature increases above $T_{C}$, the atomic character of the
system is partially restored resulting in an atomic like susceptibility with an
effective moment: 
\begin{equation}
\chi _{q=0}=\frac{\mu _{eff}^{2}}{3(T-T_{C})}  
\label{effective}
\end{equation}
The temperature dependence of the ordered  magnetic  moment below
the Curie temperature  and  the inverse
of the uniform  susceptibility  above the Curie point
are plotted  in  Fig. 3 together with the corresponding
experimental data for iron and nickel\cite{wolfarth,Vons}.
The LDA+DMFT calculations 
describes the magnetization curve and the slope of the
high-temperature Curie-Weiss susceptibility remarkably well.
The calculated values of
high-temperature magnetic moments extracted from Eq.(\ref{qzero}) are  $\mu
_{eff}=3.09$ $(1.50)\mu _{B}$ for Fe (Ni),  in  good agreement with the
experimental data $\mu _{eff}=3.13$ $(1.62)\mu _{B}$ for Fe (Ni)\cite{wolfarth}.

We have estimated the values of
the  Curie temperatures of Fe and Ni from the
disappearance of spin polarization in the self-consistent solution of DMFT
problem and from the Curie-Weiss law in  (Eq.(\ref{effective})).
Our estimates $T_{C}=1900$ $(700)K$    are in  reasonable agreement with
experimental values of $1043$ $(631)K$ for Fe (Ni) respectively\cite{wolfarth},
considering the single site nature of the DMFT approach.

Within dynamical mean field theory one can also compute 
the local spin susceptibility defined by 
\begin{equation}
\chi _{loc}=\frac{g_{s}^{2}}{3}\int\limits_{0}^{\beta }d\tau \left\langle 
{\bf S}\left( \tau \right) {\bf S}(0)\right\rangle  \label{local}
\end{equation}
where $g_{s}=2$ is the gyromagnetic ratio and
${\mathbf S}=\frac 12  \sum_{m, \sigma, \sigma^{\prime }} c^{\dagger}_{m \sigma}
{{\mathbf \sigma }}_{\sigma \sigma^{\prime }} c_{m \sigma^{\prime }}$ 
is single-site spin operator and
${\mathbf { \sigma }}=(\sigma_x,\sigma_y,\sigma_z)$ are Pauli matrices.
 It differs from the $q=0$ susceptibility, Eq.(\ref{qzero}),
by the absence of spin
polarization in the Weiss field of the impurity model. 
Eq.(\ref{local}) cannot
be probed  directly in experiments but  it is easily computed  
in DMFT-QMC. Its behavior as function of temperature, gives a very intuitive
picture of the degree of correlations in the system. In a weakly correlated
system we expect Eq.(\ref{local}) to be nearly temperature independent,
while in a strongly correlated system we expect a leading Curie-Weiss
behavior at high temperatures 
$
\chi _{local}={\mu _{loc}^{2}}/({3T+const})
$ 
where $\mu _{loc}$ is  an effective local magnetic moment. In the Heisenberg
model with spin $S$, $\mu _{loc}^{2}=S(S+1)g_{s}^{2}$ and for well-defined
local magnetic moments (e.g., for rare earth magnets) this quantity should
be temperature independent. For the itinerant electron magnets $\mu _{loc}$
is temperature-dependent, due to a variety of competing many body effects
such as Kondo screening, the induction of local magnetic moment
by temperature \cite{moriya} and thermal fluctuations which disorders the
moments \cite{UFN}. All these effects are included in the DMFT calculations.
The $\tau $-dependence of the correlation function $\left\langle {\bf S}%
\left( \tau \right) {\bf S}(0)\right\rangle $ results in the temperature
dependence of $\mu _{loc}$ and  is displayed  in the inserts on the Figs.1,2. Iron
can be considered as a magnet with very well-defined local moments above 
$T_{C}$ (the $\tau $-dependence of the correlation function is relatively
weak), whereas nickel is more itinerant electron magnet ( stronger $\tau $ 
-dependence of the local spin-spin autocorrelation function ).

\begin{figure}[tbp]
\centerline{\epsfig{file=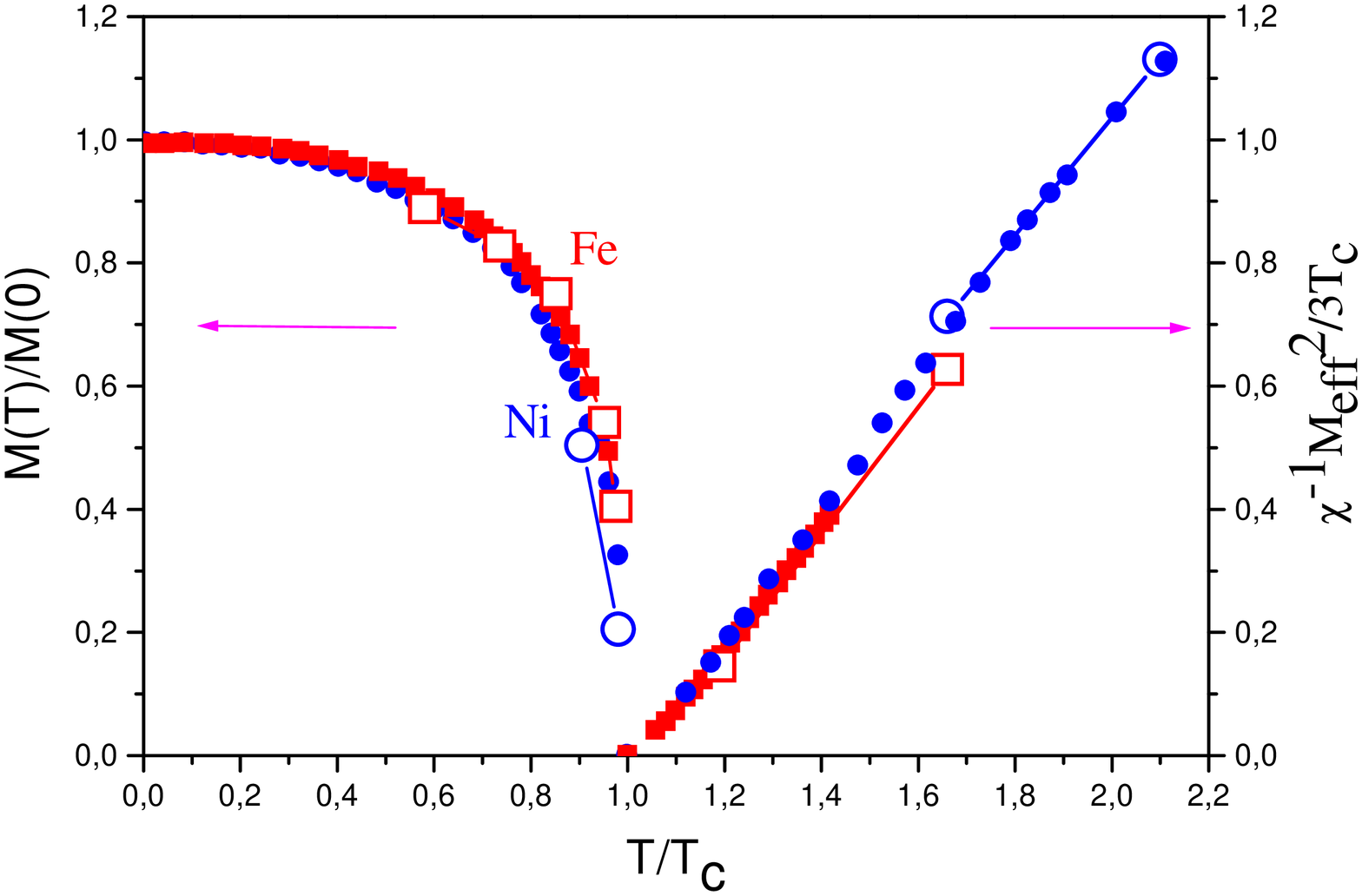, width=8cm,height=6cm,angle=0} }
\caption{Temperature  dependence of ordered  moment and the
inverse ferromagnetic suscpetibility for Fe (open square) and Ni (open
circle) compared  with experimental results for Fe (square) and Ni
(circle) (from Ref.[1,28]). }
\label{Mchi}
\end{figure}
The comparison of the values of the local and the $q=0$ susceptibility gives
a crude measure of the degree of short range order which is present above $%
T_{C}$. As expected, the moments extracted from the local susceptibility
Eq.(\ref{local}) are a bit smaller ( 2.8 \ $\mu _{B}$ for iron and 1.3 \ 
$\mu _{B}$ for nickel) than those extracted from the uniform mangetic
susceptibility. This reflects the small degree of the short-range
correlations which remain well above $T_{C}$\cite{SRO}.
The high-temperature LDA+DMFT  clearly
show the presence of a  local-moment above $T_{C}$. 
This moment, is correlated with the presence of high energy features
(of the order of the Coulomb energies) in the photomeission.
This is also true below $T_{C}$, where the spin dependence of the spectra is 
more pronounced for the satellite rigion in nickel than for   
that the quasiparticle bands near the Fermi level (Fig. 2).
This can explain the apparent discrepancies between
different experimental determinations of the high-temperature
magnetic splittings \cite{LMM,sinkovic,aebi}
as being the result of probing different energy regions. The resonant
photoemission experiments\cite{sinkovic} reflects  the presence of 
local-moment polarization
in the high-energy spectrum above Curie temperature in nickel, while the
low-energy ARPES investigations\cite{aebi} results in non-magnetic bands
near the Fermi level. This is exactly the DMFT view on the electronic
structure of transition metals above $T_{C}$. Fluctuating moments and atomic-like configurations  are large
at short times, which results in correlation effects in the high-energy
spectra such as spin-multiplet splittings. The moment is reduced at longer time
scales, corresponding to a more band-like, less correlated electronic
structure near the Fermi level.

To conclude, we presented the first results of {\it ab initio} LDA+DMFT
calculations of finite-temperature magnetic properties for Fe and Ni and
showed that many body effects which incorporate the local atomic character
of the electrons and which are ignored in the standard LDA based scheme are
essential for a simultaneous description of the magnetic properties and the
one electron spectra of itinerant electron magnets. DMFT gives a
satisfactory semiquantitative description of the physical propertis of Fe
and Ni, far from the Curie point, indicating that the critical fluctuations,
which are not included in the DMFT approximation, do not play a crucial
role, except for the immediate vicinity of the transition, and many aspects
of the physics of this system can be understood within an approach which
stresses local physics.
It would be interesting to extend this study to other itinerant magnetic
systems  with more atoms per unit cell,
such as SrRuO$_3$ which is also well described by band theory
at very low temperatures but has anomalous properties above its Curie point
\cite{schlessinger}.

ACKNOWLEDGEMENT This work is supported by the Netherlands Organization for
Scientific Research, grant NWO 047-008-16. GK was supported by the ONR,
grant No. 4-2650. A.I.L. is grateful to the Center for Materials Theory at
Rutgers University for its hospitality, during the initial stages of this
work.

\end{multicols}

\end{document}